\documentclass[aps,prl,floats,superscriptaddress,floatfix]{revtex4}
\usepackage{epsf,graphicx}
\usepackage{amssymb}
\usepackage{amsmath}
\usepackage{eepic}
\usepackage[dvips]{color}
\begin{document}

\title{Nearly-degenerate $p_x+ip_y$ and $d_{x^2-y^2}$ pairing symmetry\\
in the heavy fermion superconductor YbRh$_2$Si$_2$ }

\author{Yu Li}
\affiliation{Beijing National Laboratory for Condensed Matter Physics and
Institute of Physics, Chinese Academy of Sciences, Beijing 100190, China}
\author{Qianqian Wang}
\affiliation{Beijing National Laboratory for Condensed Matter Physics and
Institute of Physics, Chinese Academy of Sciences, Beijing 100190, China}
\affiliation{Department of Physics, Engineering Research Center for Nanophotonics and Advanced Instrument, East China Normal University, Shanghai 200062, China}
\author{Yuanji Xu}
\affiliation{Beijing National Laboratory for Condensed Matter Physics and
Institute of Physics, Chinese Academy of Sciences, Beijing 100190, China}
\affiliation{University of Chinese Academy of Sciences, Beijing 100049, China}
\author{Wenhui Xie}
\affiliation{Department of Physics, Engineering Research Center for Nanophotonics and Advanced Instrument, East China Normal University, Shanghai 200062, China}
\author{Yi-feng Yang}
\email[]{yifeng@iphy.ac.cn}
\affiliation{Beijing National Laboratory for Condensed Matter Physics and
Institute of Physics, Chinese Academy of Sciences, Beijing 100190, China}
\affiliation{University of Chinese Academy of Sciences, Beijing 100049, China}
\affiliation{Songshan Lake Materials Laboratory, Dongguan, Guangdong 523808, China}

\date{\today}

\begin{abstract}
Recent discovery of superconductivity in YbRh$_2$Si$_2$ has raised particular interest in its pairing mechanism and gap symmetry. Here we propose a phenomenological theory of its superconductivity and investigate possible gap structures by solving the multiband Eliashberg equations combining realistic Fermi surfaces from first-principles calculations and a quantum critical form of magnetic pairing interactions. The resulting gap symmetry shows sensitive dependence on the in-plane propagation wave vector of the quantum critical fluctuations, suggesting that superconductivity in YbRh$_2$Si$_2$ is located on the border of $(p_x+ip_y)$ and $d_{x^2-y^2}$-wave solutions. This leads to two candidate phase diagrams: one has only a spin-triplet $(p_x+ip_y)$-wave superconducting phase; the other contains multiple phases with a spin-singlet $d_{x^2-y^2}$-wave state at zero field and a field-induced spin-triplet $(p_x+ip_y)$-wave state. In addition, the electron pairing is found to be dominated by the `jungle-gym' Fermi surface rather than the `doughnut'-like one, in contrast to previous thought. This requests a more elaborate and renewed understanding of the electronic properties of YbRh$_2$Si$_2$.
\end{abstract}

\maketitle

Recent discovery of superconductivity below 2 mK in YbRh$_2$Si$_2$ has doubled the total number of Yb-based heavy fermion superconductors  \cite{Schuberth2016}. While YbRh$_2$Si$_2$ has been a subject of decade-long studies due to its peculiar quantum critical properties \cite{Custers2003,Paschen2004,Friedemann2009,Stockert2011}, this latest discovery has stimulated new interest concerning the nature of its pairing symmetry. At higher temperatures, the angle-resolved photoemission spectroscopy (ARPES) has observed large Fermi surfaces of dominant $f$-orbital characters down to 1 K~\cite{Kummer2015}, implying the existence of itinerant Yb-4$f$ electrons for superconducting pairing. Indeed, it is currently believed that superconductivity in YbRh$_2$Si$_2$ is formed of heavy-electron pairs. Still, question remains concerning the origin of potential pairing glues and symmetry of the gap structure. A satisfactory understanding of the pairing mechanism is still lacking.

A probable candidate for the pairing glue might come from magnetic quantum critical fluctuations. Although superconductivity was so far only explored in the antiferromagnetic (AFM) phase below $T_{\text{N}}=70$~mK \cite{Trovarelli2000}, it is close to the quantum critical point due to the small critical field (0.06$\,$T along the $a$-$b$ plane and 0.66$\,$T along the $c$-axis) and its microscopic coexistence with AFM has been excluded \cite{Schuberth2016}. The magnetically ordered phase is believed to contain significant fluctuations. It has a tiny ordered moment ($< 0.1\mu_{\text{B}}$/Yb$^{3+}$) compared to the effective moment, $\mu_{\text{eff}}\approx 1.4\mu_{\text{B}}$/Yb$^{3+}$, derived from a Curie-Weiss fit of the susceptibility slightly above $T_{\text{N}}$ \cite{Trovarelli2000,Gegenwart2002}. Nuclear magnetic resonance has revealed strong AFM fluctuations near the quantum critical point (QCP) \cite{Ishida2002}. By contrast, neutron scattering experiments have detected significant ferromagnetic (FM) fluctuations below 30 K, which evolve into incommensurate in-plane AFM correlations with a propagation wave vector $\mathbf{Q}_\perp=\pm (0.14\pm 0.04,0.14\pm 0.04)$ at 0.1 K \cite{Stock2012}. Thus, superconductivity in YbRh$_2$Si$_2$ might also be mediated by magnetic quantum critical fluctuations, similar to many other heavy fermion superconductors including CeCu$_2$Si$_2$, CeRhIn$_5$, UGe$_2$, etc., in which superconductivity can also be present within a magnetic phase but mediated by spin fluctuations \cite{Pfleiderer2009,White2015,Scalapino2012,Yang2015}.

From theoretical perspective, the phase-separated coexistence of a long-range magnetic order should play no major role in determining the superconducting gap symmetry. For simplicity, one might ignore first the presence of antiferromagnetism and consider in theory solely the superconducting instability. This allows us to calculate the pairing symmetry based on realistic heavy electron band structures derived from first-principles calculations and a phenomenological form of magnetic quantum critical pairing interactions. We find that YbRh$_2$Si$_2$ is located on the border of a $d_{x^2-y^2}$-wave spin-singlet state and a $(p_x+ip_y)$-wave spin-triplet state. The exact ground state depends sensitively on the in-plane ($h$) component of the vector $\mathbf{Q}\equiv(h,h,l)$ of the pairing interactions. This yields two candidate scenarios: one with spin-triplet $(p_x+ip_y)$-wave pairing, and the other with a spin-singlet $d_{x^2-y^2}$-wave state at zero field and an induced spin-triplet $(p_x+ip_y)$-wave state at high field.

\begin{figure}[t]
\centering\includegraphics[width=0.48\textwidth]{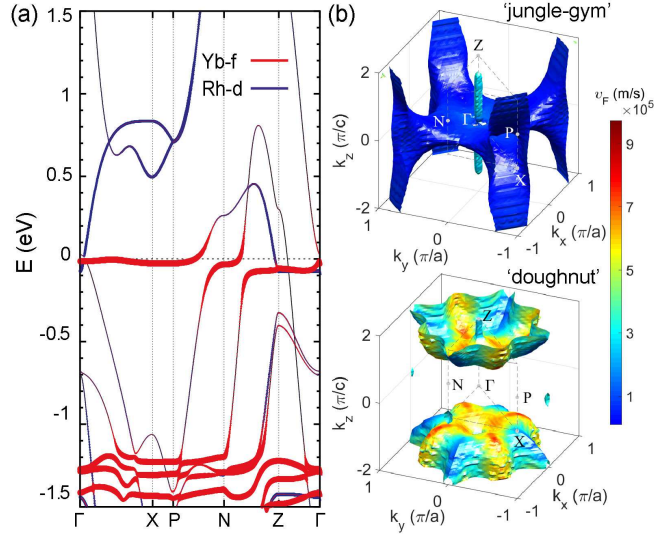}
\caption{(a) Electronic band structures of YbRh$_2$Si$_2$ from first-principles calculations, showing the $f$-electron character of the flat hybridization bands near the Fermi energy. (b) Illustration of the `jungle-gym' electron Fermi surface and the `doughnut'-like hole Fermi surface. The color represents the corresponding Fermi velocity, where the renormalization effect due to $Z_\mu$ is not  included.}
\label{fig1}
\end{figure}

The electronic structures of YbRh$_2$Si$_2$ were obtained using the density functional theory (DFT) taking into consideration both the spin-orbit coupling and an effective Coulomb interaction $U=8\,$eV \cite{Perdew1996,Anisimov1997,Suzuki2010,Blaha2018}. As shown in Fig. \ref{fig1}, we find two flat bands that cross the Fermi energy and exhibit strong hybridization between Yb-4$f$ and Rh-4$d$ orbitals. The electron band along the $\Gamma$-X-P path produces the so-called `jungle-gym' electron Fermi surface \cite{Wigger2007}, and the hole band around Z point yields the `doughnut'-like hole Fermi surface. The results are plotted in Fig. \ref{fig1}(b) and the value of $U$ was chosen to yield the same topological structures as in previous calculations~\cite{Friedemann2010,Zwicknagl2016}. Experimentally, the `doughnut'-like hole Fermi surface has been observed by ARPES \cite{Wigger2007,Vyalikh2008,Vyalikh2009,Vyalikh2010,Danzenbacher2011,Mo2012, Kummer2012,Kummer2015}, in agreement with theoretical predictions \cite{Friedemann2010,Zwicknagl2016}, while the `jungle-gym' electron Fermi surface was missing but argued to be covered up by surface states {\cite{Kummer2015}}. In de Haas-van Alphen (dHvA) measurements \cite{Rourke2008,Sutton2010}, a high-frequency mode has been detected and attributed to the `jungle-gym' Fermi surface. More detailed comparisons on the mass enhancement can be found in Supplemental Materials \cite{Supp}. The agreement suggests that DFT+$U$  provides a reasonable starting point for superconducting calculations of YbRh$_2$Si$_2$.

The renormalization effect of quantum critical interactions and the pairing symmetry can be investigated by solving the linearized Eliashberg equations \cite{Monthoux1992,Nishiyama2013,Yang2014,Li2018},
\begin{eqnarray}
Z_{\mu}\left(\bold{k},i\omega_n \right)&=&1+\frac{\pi T}{\omega_n}\sum_{\nu,m} \oint_{\text{FS}_\nu}
\frac{d\bold{k^{\prime}_{\parallel}}}{(2\pi)^3v_{\nu,\bold{k}^{\prime}_\text{F}}}\text{sgn} \left( \omega_m \right) \nonumber \\
&\times& V^{\mu\nu}\left( \bold{k-k^{\prime}},i\omega_n-i\omega_m \right),\nonumber\\
\lambda\phi_{\mu}\left(\bold{k},i\omega_n \right)&=&-C\pi T\sum_{\nu,m}
\oint_{\text{FS}_\nu}\frac{d\bold{k^{\prime}_{\parallel}}}{(2\pi)^3v_{\nu,\bold{k}^{\prime}_\text{F}}}  \\
&\times& \frac{V^{\mu\nu}\left(\bold{k-k^{\prime}},i\omega_n-i\omega_m \right)} {\left| \omega_m
Z_{\nu}\left(\bold{k^{\prime}},i\omega_m \right)\right|}\phi_{\nu} \left(\bold{k^{\prime}},i\omega_m \right),\nonumber
\label{eq1}
\end{eqnarray}
where $\mu$ and $\nu$ are the band indices, FS$_\nu$ denotes the integral over the Fermi surface of band $\nu$, $v_{\nu,\bold{k}^{\prime}_\text{F}}$ is the corresponding Fermi velocity, $V^{\mu\nu}$ is the intraband ($\mu=\nu$) or interband ($\mu\neq\nu$) interactions, $\omega_{n/m}$ is the fermionic Matsubara frequency, $Z_{\mu}$ is the renormalization function, and $\phi_{\mu}$ is the anomalous self-energy related to the gap function, $\Delta_{\mu}=\phi_{\mu}/Z_{\mu}$. It is important to note that $Z_\mu$ might not only provide the major mass enhancement entering the quantum critical regime \cite{Gegenwart2006}, but also reduces the spectral weight of pairing quasiparticles. Thus it would be incorrect to start with fully renormalized bands for superconducting calculations \cite{Supp}. The prefactor $C$ is unity for spin-singlet pairing and $-1/3$ for spin-triplet pairing. $\lambda$ is the eigenvalue of the kernel matrix for each pairing channel and its largest value determines the dominant pairing state at $T_c$. Unlike iron-pnictides, where the Fermi surfaces are mostly quasi-two-dimensional and nearly isotropic, the Fermi surfaces here are highly anisotropic and three-dimensional, so the superconducting gap structures cannot be easily captured by the low-order trigonometric harmonics near the high-symmetric points \cite{Maiti2011,Chubukov2012}. It is therefore necessary to derive the detailed gap structures by solving the Eliashberg equations numerically.

However, there are still two obstacles before we can proceed to do the calculations. First, controversy still remains regarding the exact form of the magnetic quantum critical fluctuations. While different theories have been proposed based on local quantum criticality \cite{Si2001,Si2014} or critical quasiparticles \cite{Wolfle2011,Abrahams2012,Wolfle2017}, neutron scattering experiments seem to have detected simple spin-density-wave (SDW) type fluctuations \cite{Stock2012}. We will not try to judge these different scenarios. Rather, we adopt a generic and phenomenological form for the pairing interactions \cite{Millis1990,Monthoux1991,Monthoux1992,Nishiyama2013,Yang2014,Li2018},
\begin{equation}
V^{\mu\nu}(\mathbf{q},i\nu_n)=\frac{V_0^{\mu\nu}}{1+\xi^2\left(\mathbf{q}- \mathbf{Q}\right)^2+\left|\nu_n/\Lambda_{\text{sf}}\right|^{\alpha}},
\label{eq2}
\end{equation}
where $V_0^{\mu\nu}$ are free parameters controlling the relative strength of intra- and interband pairing forces. The exponent $\alpha$ defines different quantum critical scenarios and takes the value of 1 for SDW \cite{Stock2012}, 0.75 for local quantum criticality \cite{Si2001,Si2014} and 0.5 for critical quasiparticle theory \cite{Wolfle2011,Abrahams2012,Wolfle2017}. We estimated the correlation length $\xi\approx 6\,\mathring{\text{A}}$ very crudely from neutron scattering experiments \cite{Stock2012} and chose the characteristic spin-fluctuation frequency $\Lambda_{\text{sf}}\approx 1\,$meV such that the magnetic Fermi energy $\Gamma_{\text{sf}}=\Lambda_{\text{sf}}(\xi/a)^2\approx 2.2\,$meV equals roughly the Kondo energy scale \cite{Schuberth2016}. For numerical calculations, we discretize the whole Brillouin zone into 70$\times$70$\times$70 $\bold{k}$-meshes and take 8192 Matsubara frequencies for the $\omega_n$-summation to be cut off at around $\Gamma_{\text{sf}}$. The gap structure in the momentum space is then solved with the approximation $g_{\mu,\bold{k}}\equiv\Delta_\mu(\bold{k},i\omega_n)\approx \Delta_\mu(\bold{k},i\pi T_c)$. Interestingly, our calculations show that the gap symmetry is independent of $\alpha$ but mainly determined by the momentum structure of the pairing interactions. Here comes the second obstacle that concerns $\mathbf{Q}=(h,h,l)$. Experimentally, it evolves with temperature from $h=l=0$ (FM) below 30$\,$K to $h=0.14\pm 0.04$ (AFM) at 0.1$\,$K \cite{Stock2012}. Since its exact value for the electron pairing at $T_c$ is yet to be measured, we are forced to consider a wide range of possibilities around these experimental observations. Such a strategy turns out to be helpful and reveals the nearly degenerate nature of the superconductivity in YbRh$_2$Si$_2$.

\begin{figure}[t]
\centering\includegraphics[width=0.48\textwidth]{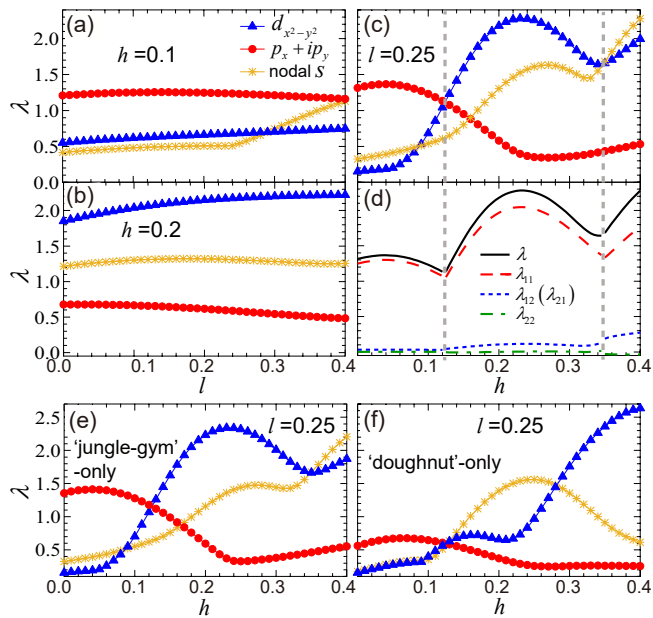}
\caption{Evolution of three key eigenvalues $\lambda$ with varying $\mathbf{Q}=(h,h,l)$ for (a) $h=0.1$; (b) $h=0.2$; (c)  $l=0.25$. (d) Band-resolved eigenvalues for the leading solution in (c) as a function of $h$. (e) and (f) plot $\lambda$ as a function of $h$ with given $l=0.25$ in the one-band calculations for each of the two Fermi surfaces. For clarity, eigenvalues that never dominate are not shown in all panels.}
\label{fig2}
\end{figure}

\begin{figure}[t]
\centering\includegraphics[width=0.48\textwidth]{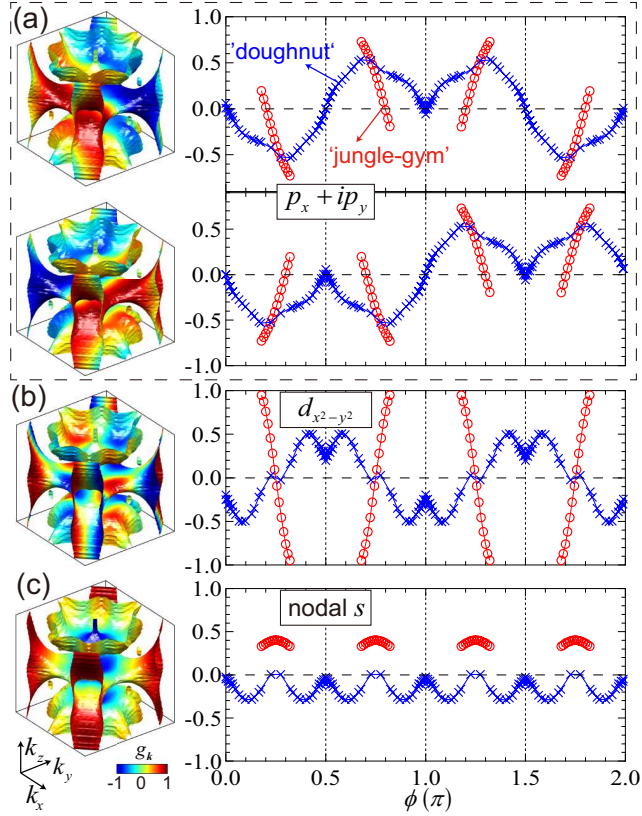}
\caption{Distribution of typical gap structures on the Fermi surfaces and with the azimuthal angle $\phi$ for (a) the $p_x$ and $p_y$ components of the leading $p_x+ip_y$-wave solution for $\mathbf{Q}=(0.1,0.1,0.25)$; (b) the leading $d_{x^2-y^2}$-wave solution for $\mathbf{Q}=(0.2,0.2,0.25)$; (c) the leading nodal $s$-wave solution for $\mathbf{Q}=(0.4,0.4,0.25)$. The results are shown for $k_z=1.5\pi/c$ plane.}
\label{fig3}
\end{figure}

Figure~\ref{fig2} plots the eigenvalues of three major pairing channels for different choices of $\mathbf{Q}$. For simplicity, we only present the data for $\alpha=1$ and assume a band-independent $V_0^{\mu\nu}$. We have examined other choices in a reasonable range of variations and found no qualitative influence on our main conclusions (see Supplemental Materials~\cite{Supp}). Figures~\ref{fig2}(a) and \ref{fig2}(b) compare the eigenvalues as a function of $l$ for fixed $h=0.1$ and 0.2, revealing a leading solution of either $(p_x+ip_y)$ or $d_{x^2-y^2}$-wave over a wide parameter range of $l$. Thus the electron pairing is insensitive to magnetic fluctuations along $c$-axis. We also plot the $h$-dependence of the eigenvalues for a typical $l=0.25$ in Fig.~\ref{fig2}(c), where we could see clear transitions of the leading pairing channel from $(p_x+ip_y)$ to $d_{x^2-y^2}$ at $h\approx 0.13$ and then to a nodal $s$-wave solution at $h\approx0.35$, indicating that in-plane magnetic fluctuations play a crucial role in determining the pairing symmetry. For clarity, typical gap structures of above solutions are plotted in Fig.~\ref{fig3} for different values of $h$ at fixed $l=0.25$. For $h=0.1$, we derive a two-fold degenerate solution with $p_x$ and $p_y$ symmetry as shown in their dependence on the azimuthal angle ($\phi$). Their mixture gives the chiral $(p_x+ip_y)$-wave gap to minimize the pairing energy, $E=-\frac13\sum_{\substack{\bold{k},\bold{k}^{\prime},\mu,\\ \nu,\alpha,\beta}} V^{\mu\nu}_{\bold{k}\bold{k}^{\prime}}\langle c^{\dagger}_{\mu,\bold{k}\alpha}c^{\dagger}_{\mu,-\bold{k}\beta}\rangle\langle c_{\nu,-\bold{k}^{\prime}\beta}c_{\nu,\bold{k}^{\prime}\alpha}\rangle$, where $\alpha$ and $\beta$ are spin indices. For $h=0.2$, a $d_{x^2-y^2}$-wave gap is obtained which changes sign when $\phi$ rotates by $\pi/2$ and contains nodes on the $k_x=\pm k_y$ plane. For $h=0.4$, we identify a nodal $s$-wave solution with accidental nodes on the `doughnut'-like Fermi surface.

To extract key factors that determine the pairing symmetry, we separate out contributions from each Fermi surface and define the band-resolved eigenvalues~\cite{Maier2009},
\begin{equation}
\lambda_{\mu\nu}=\frac{\oint_{\text{FS}_\mu}\frac{d\bold{k_{\parallel}}} {(2\pi)^3v_{\mu,\bold{k}_\text{F}}}\oint_{\text{FS}_\nu}\frac{d\bold{k^{\prime}_{\parallel}}} {(2\pi)^3v_{\nu,\bold{k}^{\prime}_\text{F}}}K^{\mu\nu}_{\bold{k},\bold{k^{\prime}}} g^{*}_{\mu,\bold{k}} g_{\nu,\bold{k^{\prime}}}}{\oint_ {\text{FS}_\mu}\frac{d\bold{k_{\parallel}}} {(2\pi)^3v_{\mu,\bold{k}_\text{F}}}\left|g_{\mu,\bold{k}} \right|^2},
\end{equation}
where $K^{\mu\nu}_{\bold{k},\bold{k^{\prime}}}=-C\pi T_c\sum_{m} V^{\mu\nu}_{\bold{k},\bold{k^{\prime}}}(i\pi T_c-i\omega_m)/\left| \omega_m \right|$ and $V^{\mu\nu}_{\bold{k},\bold{k^{\prime}}}(i\nu_n)=\left[ V^{\mu\nu}(\bold{k}-\bold{k^{\prime}},i\nu_n)\pm V^{\mu\nu}(\bold{k}+\bold{k^{\prime}},i\nu_n)\right]/2$ for spin-singlet ($+$) and triplet ($-$) pairings, respectively. $\lambda_{\mu\nu}$ represents the effective pairing strength between the $\mu$ and $\nu$ Fermi surfaces. For $\mu=\nu$, it denotes the intraband contribution within each Fermi surface, while for $\mu\neq\nu$, it accounts for the contribution from interband pair scattering. The true eigenvalue is a sum of all terms, $\lambda=\sum_{\mu,\nu}\lambda_{\mu\nu}$. Figure \ref{fig2}(d) plots the band-resolved $\lambda_{\mu\nu}$ for the leading solutions in each regime as a function of $h$. In all three regimes, $\lambda_{11}$ is always the largest, implying that the `jungle-gym' electron Fermi surface is the major player in forming superconductivity. To understand this, we consider the electron pairing on each single Fermi surface alone and solve the one band Eliashberg equations with the same parameters. The results are compared in Figs.~\ref{fig2}(e) and \ref{fig2}(f). For small $h$, both Fermi surfaces have the same leading $(p_x+ip_y)$-wave solution owing to the ferromagnetic-like pairing interaction; while for intermediate $h$, the `jungle-gym' Fermi surface favors a $d_{x^2-y^2}$-wave gap but the `doughnut'-like Fermi surface yields a nodal $s$-wave gap. Thus for the two-band model, the `jungle-gym' Fermi surface dominates the leading pairing channel and gives rise to the $d_{x^2-y^2}$-wave gap for intermediate $h$. We attribute this to the special topology of the `jungle-gym' Fermi surface which is more strongly nested and matches better the momentum structure of the pairing glue than the `doughnut'-like one (see Supplemental Materials for an illustration of their respective nesting properties~\cite{Supp}). The fact that $\lambda_{22}$ is suppressed to almost zero in the two-band calculations compared to its value in the single-band calculations reflects microscopic competition of the pair formation on two Fermi surfaces. We would like to note that the `doughnut'-like Fermi surface was often treated as the major or only player in previous literatures. Our results suggest that this might be an oversimplified picture.

\begin{figure}[t]
\centering\includegraphics[width=0.48\textwidth]{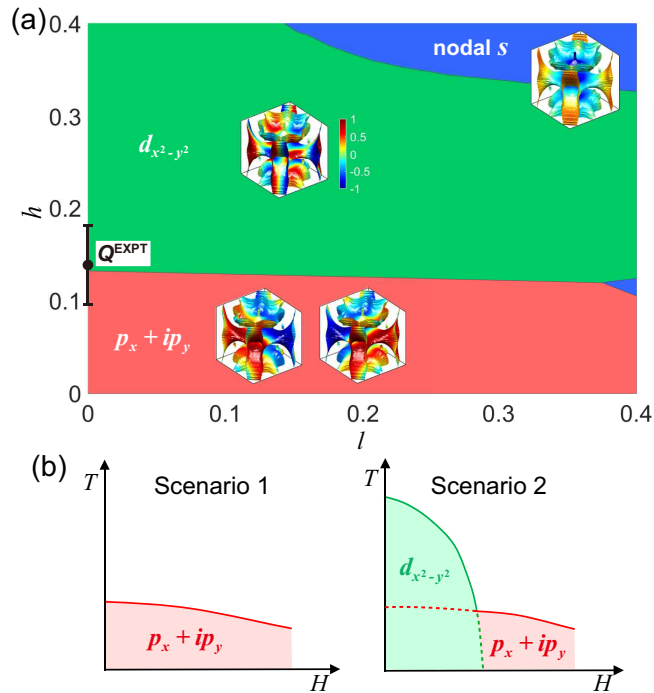}
\caption{(a) Theoretical phase diagram of the superconductivity in YbRh$_2$Si$_2$ as a function of the propagation wave vector $\mathbf{Q}=(h,h,l)$ of pairing interactions. The insets illustrate the gap structures in each phase. $\mathbf{Q}^{\text{EXPT}}=(0.14,0.14,0)$ marks the observed $\mathbf{Q}$ in neutron scattering experiments at 0.1 K. The error bar represents the experimental error, $h=0.14\pm 0.04$. (b) Two candidate $T$-$H$ phase diagrams of the superconductivity with dominant $(p_x+ip_y)$ or $d_{x^2-y^2}$-wave solutions at zero magnetic field.}
\label{fig4}
\end{figure}

Figure~\ref{fig4} summarizes all the leading solutions on a global phase diagram of the superconductivity with varying $\mathbf{Q}$ for YbRh$_2$Si$_2$. Among them, $(p_x+i p_y)$ dominates the lower part of the phase diagram with small $h$, $d_{x^2-y^2}$ governs most of the upper part, while the nodal $s$-wave solution only occurs at the corners. These are not unexpected, as the $(p_x+ip_y)$-wave solution is a spin-triplet state favored by FM-like fluctuations with small $h$, $d_{x^2-y^2}$ originates from the nested `jungle-gym' Fermi surface and associated AFM fluctuations, and the nodal $s$-wave solution, which is not crucial, might appear when large-momentum transfers start to correlate Cooper pairs on different portions of the Fermi surfaces. The true ground state of the superconductivity in YbRh$_2$Si$_2$ can then be determined if the exact wave vector responsible for the pairing below $T_c$ are known. Unfortunately, this  requires very challenging experiment which so far has not yet been done. For candidate $\mathbf{Q}^{\text{EXPT}}=(0.14,0.14,0)$ measured by neutron scattering at 0.1 K above the AFM order \cite{Stock2012}, a $d_{x^2-y^2}$-wave gap is obtained but located very close to the $d_{x^2-y^2}$ and $(p_x+ip_y)$ phase boundary. A slight variation due to experimental error ($h=0.14\pm 0.04$) would lead to a spin-triplet $(p_x+ip_y)$-wave pairing. Further uncertainty may arise from potential temperature evolution of the $\mathbf{Q}$-vector. Very recently, it was also proposed in the critical quasiparticle theory that additional energy fluctuations might favor a $p$-wave solution \cite{Kang2018}. Thus, a natural statement would be that the superconductivity in YbRh$_2$Si$_2$ is located in a delicate position with nearly-degenerate $d_{x^2-y^2}$ and $(p_x+ip_y)$-wave symmetries. It is easy to imagine that a magnetic field would presumably shift the balance and promote the $(p_x+ip_y)$-wave spin-triplet solution. We thus speculate two possible scenarios for the $T$-$H$ (temperature-magnetic field) phase diagrams as sketched schematically in Fig.~\ref{fig4}(b). If the $(p_x+ip_y)$-wave spin-triplet state wins out, there would only be a single superconducting phase under field. By contrast, if the $d_{x^2-y^2}$-wave spin-singlet state is stronger, it might be more rapidly suppressed by external magnetic field and the $(p_x+ip_y)$-wave spin-triplet state could then be induced, causing multiple superconducting phases.

Yet experiments so far are inconclusive. In the original work, only one superconducting phase was reported below about 2 mK \cite{Schuberth2016}. It has an extrapolated upper critical field, $H_{c2}(T\rightarrow0)\approx 30-50\,$mT, comparable to its orbital limiting field, $H_{c2,{\text{orb}}}=0.693(-dH_{c2}/dT)\vert_{T_c}T_c\approx 35\,$mT \cite{Werthamer1966} but well beyond the Pauli limiting field, $H_{c2,\text{P}}=1.84T_c\approx 3.7\,$mT \cite{Clogston1962,Chandrasekhar1962}. Since the Pauli limit is generally associated with pair breaking of the spin-singlet, the fact that $H_{c2,\text{P}}\ll H_{c2,\text{orb}}\approx H_{c2}$ manifests dominant orbital effects and suggests that this single superconducting phase should be of spin-triplet pairing, in agreement with the first scenario in Fig.~\ref{fig4}(b). However, latest experiment reported a different zero-field superconducting phase with $T_c\approx 6\,$mK and its transition to a field-induced phase with $T_c\approx 2\,$mK at about 4 mT \cite{Saunders2018}, pointing towards the possibility of multiple superconducting phases tuned by the magnetic field. The two phases show very different field dependence of $T_c$. While the field-induced phase is very similar to the originally observed (spin-triplet) one \cite{Schuberth2016}, the zero-field phase has an extrapolated upper critical field, $H_{c2}(T\rightarrow0)\approx 4\,$mT, which is below its Pauli limiting field, $H_{c2,{\text{P}}}=1.84T_c\approx 11\,$mT. Since $H_{c2}<H_{c2,\text{P}}$, the zero-field phase is most probably spin-singlet. Thus the latest experiment seems to support the second scenario proposed in Fig.~\ref{fig4}(b). If this is the case, our theory predicts that the zero-field phase should be a $d_{x^2-y^2}$-wave spin-singlet state, and the field-induced phase would then be a $(p_x+ip_y)$-wave spin-triplet state. This implies the existence of multiple superconducting phases is an intrinsic electronic property of YbRh$_2$Si$_2$, although the presence of nuclear order might play a role in the phase diagram. The seeming ``inconsistency" of two experiments, possibly influenced by some yet-to-be-identified factors in the experimental setup, might actually be a supporting evidence for our proposal of two nearly-degenerate pairing states.

To summarize, we have proposed a quantum critical pairing mechanism for the newly-discovered superconductivity in YbRh$_2$Si$_2$ and explored its possible gap symmetry using phenomenological pairing interactions with realistic band structures from first-principles calculations. For proper experimental parameters, we obtain nearly-degenerate $d_{x^2-y^2}$ and $(p_x+ip_y)$-wave solutions. This leads to two candidate temperature-magnetic field phase diagrams. While the original experiment seems to support a single $(p_x+ip_y)$-wave superconducting phase, the latest experiment supports the scenario of two superconducting phases. In the latter case, our result implies a spin-singlet $d_{x^2-y^2}$-wave pairing state at zero field and a field-induced spin-triplet $(p_x+ip_y)$-wave state. Our calculations show that the `jungle-gym' Fermi surface plays the major role for electron pairing rather than the `doughnut'-like one. This differs from the conventional picture and requests more elaborate investigations in pursuit of a concrete and thorough understanding of the electronic properties of YbRh$_2$Si$_2$.

This work was supported by the National Natural Science Foundation of China (NSFC Grant Nos. 11774401, 11522435, 51572086), the National Key R\&D Program of China (Grant No. 2017YFA0303103), and the Youth Innovation Promotion Association of CAS.

\newpage

\subsection{Supplemental Materials}

We discuss three major aspects of our theory: (1) the rationality of DFT+$U$ band structures; (2) the effect of quantum critical renormalization in Eliashberg equations; (3) the robustness of pairing symmetry with reasonable variations of the parameters. We distinguish the hybridization and renormalization effects in superconducting calculations and end with a brief remark on the DFT+$U$+QC framework for heavy fermion studies.

\subsection{I. Comparison of our calculated Fermi surfaces with experiments}

As discussed in the main text, the topology of our calculated Fermi surfaces for YbRh$_{2}$Si$_{2}$ is consistent with previous first-principles calculations  \cite{Friedemann2010,Rourke2008} and ARPES and dHvA experiments \cite{Rourke2008,Sutton2010,Kummer2015}. Here we explore more details on the quasiparticle effective mass. Experimentally, only dHvA measurements have provided some information on the effective mass of the two Fermi surfaces. For the `jungle-gym' Fermi surface, we have $m_{1}^{\ast}/m_e\approx21\pm2$, where $m_e$ is the free electron mass. The `doughnut'-like Fermi surface exhibits a number of different modes whose masses $m_{2}^{\ast}/m_e$ vary from 5 to 13 \cite{Rourke2008,Sutton2010}. This leads to a mass ratio, $m_{1}^{\ast}/m_{2}^{\ast}\approx1.6-4.6$ between two Fermi surfaces. Moreover, the largest mass enhancement is about $m_{2}^{\ast}/m_{b}\approx14\pm1$ on the `doughnut'-like Fermi surface, compared to the band mass ($m_b$) of Rh-$d$ electrons estimated from LuRh$_{2}$Si$_{2}$ \cite{Rourke2008}.

Our calculations are in good agreement with these observations. The different characters of the two Fermi surfaces reported in the dHvA experiment may be explained by their very different velocity distributions owing to different hybridization patterns. The hybridization on the `doughnut'-like Fermi surface is highly anisotropic (see lower panel of Fig.~1(b) in the main text) and can be differentiated into two parts \cite{Kummer2015}: one with strongly hybridized character and the other of nearly pure conduction character (Rh-$d$). The Fermi velocities vary drastically from $\upsilon_{2,\text{F}}\approx 8.2\times10^{4}\,$m/s for heavy electrons to $\upsilon_{c,\text{F}}\approx 9.7\times10^{5}\,$m/s for nearly unhybridized conduction electrons, giving rise to the highest enhancement $m_{2}^{\ast}/m_{b}\approx\upsilon_{c,\text{F}}/\upsilon_{2,\text{F}}\approx12$ on the `doughnut'-like Fermi surface, consistent with the dHvA measurements. In contrast, the `jungle-gym' Fermi surface (apart from the small `pillar' around $\Gamma$ to Z line) is almost uniformly hybridized with an average Fermi velocity $\upsilon_{1,\text{F}}\approx4.3\times10^{4}\,$m/s. This gives the lower boundary of the ratio $m_{1}^{\ast}/m_{2}^{\ast}\approx\upsilon_{2,\text{F}}/\upsilon_{1,\text{F}}\approx1.9$, also in reasonable agreement with the measured one between two Fermi surfaces.

However, we should note that the renormalization effect ($Z_\mu$) is not included in above comparisons. The dHvA experiments were performed under high magnetic field (8-16$\,$T) far beyond the critical field (0.66$\,$T along the $c$-axis and 0.06$\,$T along the $a$-$b$ plane) and deep inside the Fermi liquid regime, where the quantum critical effect is suppressed, as confirmed by the rapidly reduced resistivity coefficient with increasing field away from the critical point \cite{Gegenwart2002}. Thus the agreement indicates that our DFT+$U$ calculations capture well the hybridization properties of the electronic band structures in the absence of quantum critical interactions. As is in the periodic Anderson model, DFT+$U$ calculations provide the noninteracting part of the Hamiltonian with Hubbard correction.

\subsection{II. The renormalization effect of quantum critical interactions}
In our framework, the quasiparticle mass is determined by two parts: the hybridization between Yb-$f$ and Rh-$d$ bands from DFT+$U$ calculations, and the renormalization effect due to quantum critical interactions included in the Eliashberg equations. The renormalization effect plays a major role for the mass enhancement in the critical regime. For example, the specific-heat coefficient of YbRh$_2$Si$_{2}$ has been measured and extrapolated to $\gamma^{\text{EXPT}}=1.7\,$J K$^{-2}\,$mol$^{-1}$ as $T\rightarrow0$ at zero field \cite{Gegenwart2006}, while DFT+$U$ calculations only yield $\gamma^{\text{Band}}=\pi^{2}k_{\text{B}}^{2}N_{\text{F}
}/3\approx32\,$mJ K$^{-2}$ mol$^{-1}$. Hence there must be a considerable mass enhancement from quantum criticality (QC) and other interaction effects, $\gamma^{\text{EXPT}}/\gamma^{\text{Band}}=53$. Such an overall enhancement can be well accounted for by the renormalization function $Z_\mu$ without affecting the pairing symmetry. To see this, we simplify the Eliashberg equations approximately for $Z_\mu\gg 1$ in the quantum critical regime,
\begin{equation}
Z_{\mu}\left(  \mathbf{k},i\omega_{n}\right)   \approx\frac{V_{0}^{22}}{\omega_{n}}\sum_{\nu}{\oint_{\text{FS}_{\nu}}}\frac{d\mathbf{k}_{//}^{\prime}}{\left(  2\pi\right)  ^{3}\upsilon_{\nu,\mathbf{k}_\text{F}^{\prime}}}P_{\mu\nu}\left(  \mathbf{k-k^{\prime}},i\omega_{n}\right),
\end{equation}
where
\begin{equation}
P_{\mu\nu}\left(  \mathbf{k-k^{\prime}},i\omega_{n}\right)  =\pi
T\sum_{i\omega_{m}}\text{sgn}\left(  \omega_{m}\right)  \tilde{V}^{\mu\nu
}\left(  \mathbf{k-k^{\prime}},i\omega_{n}-i\omega_{m}\right)  ,
\end{equation}
and $\tilde{V}^{\mu\nu}=V^{\mu\nu}/V_0^{22}$, namely,
\begin{equation}
\tilde{V}^{\mu\nu}\left(  \mathbf{k-k^{\prime}},i\omega_{n}-i\omega_{m}\right)
=\frac{r^{\mu\nu}}{1+\xi^{2}\left(  \mathbf{k-k^{\prime}-Q}\right)
^{2}+\left\vert \omega_{n}-\omega_{m}\right\vert /\omega_{sf}},
\end{equation}
with $r^{\mu\nu}=V_0^{\mu\nu}/V_0^{22}$. Thus an overall mass enhancement can always be obtained by increasing $V_0^{22}$ with fixed $r^{\mu\nu}$. Accordingly, the eigen equation of the anomalous self-energy may also be rewritten as
\begin{equation}
\lambda\phi_{\mu}\left(  \mathbf{k},i\omega_{n}\right)   \approx-C\pi T\sum_{\nu}{\oint_{\text{FS}_{\nu}}}\frac{d\mathbf{k}_{//}^{\prime}}{\left(  2\pi\right)  ^{3}\upsilon_{\nu,\mathbf{k}_\text{F}^{\prime}}}\sum_{i\omega_{m}}\frac{\tilde{V}^{\mu\nu}\left(  \mathbf{k-k^{\prime}},i\omega_{n}-i\omega_{m}\right)  }{\left\vert\sum_{\kappa}{\oint_{\text{FS}_{\kappa}}}\frac{d\mathbf{k}_{//}^{\prime\prime}}{\left(  2\pi\right)  ^{3}\upsilon_{\kappa,\mathbf{k}_\text{F}^{\prime\prime}}}P_{\nu\kappa}\left(\mathbf{k}^{\prime}\mathbf{-k^{\prime\prime}},i\omega_{m}\right)  \right\vert}\phi_{\nu}\left(  \mathbf{k^{\prime}},i\omega_{m}\right),
\end{equation}
in which the overall factor $V_{0}^{22}$ is cancelled out. We therefore conclude that the mass enhancement due to quantum criticality can be easily accounted for by an overall scaling factor of $V_0^{\mu\nu}$ without affecting the pairing symmetry.

\begin{figure}[h]
\centering\includegraphics[width=0.6\textwidth]{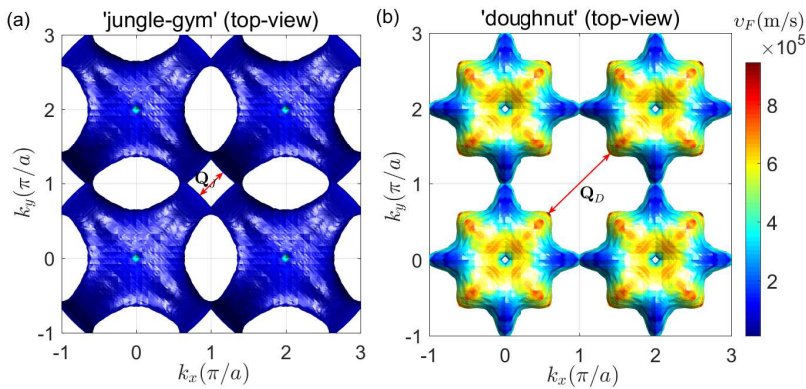}\caption{
  Illustration of the nesting properties along the (110) direction for (a) the `jungle-gym' Fermi surface and (b) the `doughnut'-like Fermi surface. The vectors labeled in the figure are $\mathbf{Q}_{J}\approx(0.16, 0.16, 0)$ and $\mathbf{Q}_{D} \approx(0.4, 0.4, 0)$.}
\label{figS1}
\end{figure}

On the other hand, the renormalization function might contribute a factor $Z_1/Z_2$ on the mass ratio between two Fermi surfaces. Our calculations yield an average $\bar{Z}_{1}/\bar{Z}_{2}\approx2.4$ in the critical regime. The overall mass ratio may then be modified to $m_{1}^{\ast}/m_{2}^{\ast}\approx\bar{Z}_{1}\upsilon_{2,\text{F}}/\bar{Z}_{2}\upsilon_{1,\text{F}}\approx4.6$, which is still within the experimental range but should be best examined in the quantum critical regime in future experiments. Such an enhancement is not arbitrary but has its root in their different nesting properties of two Fermi surfaces. As shown in Fig.~\ref{figS1}, the `jungle-gym' Fermi surface is nested with $\mathbf{Q}_{J}\approx(0.16, 0.16, 0)$, which is within the range of $\mathbf{Q}^{\text{EXPT}}=(0.14\pm0.04, 0.14\pm0.04, 0)$. A simple calculation of the Lindhard susceptibility also confirms the nesting property of the `jungle-gym' Fermi surface at the experimental wave vector compared to that of the `doughnut'-like Fermi surface. Thus the `jungle-gym'  Fermi surface is supposed to be more renormalized by quantum critical interactions.

\subsection{III. The robustness of our conclusion with varying parameters}

We have shown that our obtained mass enhancement is reasonable and consistent with current experimental observations. We further show that quantum criticality provides the major source for mass enhancement near the critical point. Both effects have already been taken into consideration in our theory. However, in the absence of an exact theory of heavy fermion physics, it is still reasonable to ask if our results are robust against possible (but small) variations of the parameters. Here we consider two possibilities: (1) the variation of $r^{\mu\nu}$; (2) the variation of the mass ratio between two Fermi surfaces.

\subsubsection{1. Variation of $r^{\mu\nu}$}
\begin{figure}[h]
\centering\includegraphics[width=0.65\textwidth]{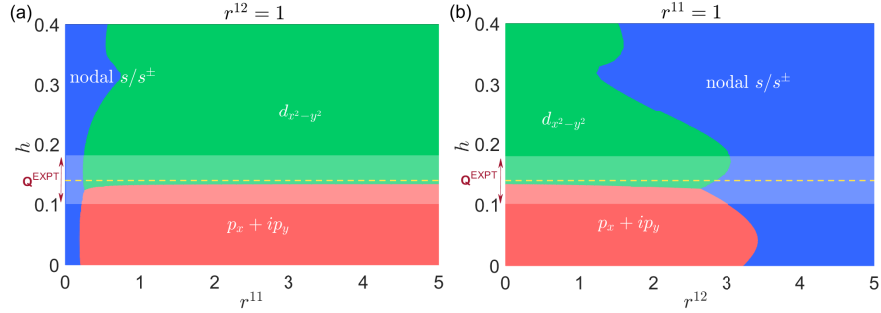}\caption{Extended phase diagrams of superconductivity with (a) $(r^{11}, h)$ and (b) $(r^{12}, h)$. The gray regions and dashed lines mark the range of $\mathbf{Q}^{\text{EXPT}} =(0.14\pm0.04, 0.14\pm0.04, 0)$.}
\label{figS2}
\end{figure}

Since the pairing symmetry is insensitive to $l$ for $\mathbf{Q}=(h, h, l)$ and not affected by an overall scaling factor of $V_0^{\mu\nu}$, we  only calculate the phase diagrams with respect to varying $h$ and $r^{11}$ or $r^{12}$. As can be seen in Fig.~\ref{figS2}, the $d_{x^2-y^2}$ and $p_x+ip_y$ ($d$-$p$) phase boundary is almost unchanged with both parameters and the pairing symmetry only deviates from the boundary when $r^{11}$ is reduced by a factor of 4 or $r^{12}$ is enhanced by a factor of 3, where the pairing becomes solely $s$-wave within experimental range of $h$, in contradiction with the presence of $p$-wave in experiments. This is a large enhancement of the parameters, considering that both Fermi surfaces originate from the same $f$ orbital in YbRh$_2$Si$_2$, and there is no reason to think differently about inter- or intra-orbital scatterings. In fact, neutron scattering intensity can be well explained by a field-induced resonance assuming a single $f$ orbital for the low-energy state \cite{Stock2012}.

\subsubsection{2. Variation of the mass ratio}
The mass ratio between two Fermi surfaces may be tuned either by the band hybridization as reflected in the Fermi velocities $\upsilon_{\mu,\mathbf{k}_\text{F}}$ given by the DFT+$U$ calculations or the quantum critical interaction included through the renormalization function $Z_{\mu}(  \mathbf{k},i\omega_{n})$ contained in the Eliashberg equations. Their variation may be seen by a band-dependent rescaling, $Z_{\mu}(  \mathbf{k},i\omega_{n})  \rightarrow\gamma_{\mu}^{Z}Z_{\mu}(  \mathbf{k},i\omega_{n})$ or $\upsilon_{\mu,\mathbf{k}_\text{F}}\rightarrow\gamma_{\mu}^{\upsilon}\upsilon_{\mu,\mathbf{k}_\text{F}}$, respectively. We discuss them separately.

(1) For $Z_{\mu}(  \mathbf{k},i\omega_{n})  \rightarrow\gamma_{\mu}^{Z}Z_{\mu}(  \mathbf{k},i\omega_{n})$, the gap equation becomes
\begin{equation}
\lambda^{\prime}\phi_{\mu}\left(  \mathbf{k},i\omega_{n}\right)  =-C\pi T\sum_{\nu}{\oint_{\text{FS}_{\nu}}}\frac{d\mathbf{k}_{//}^{\prime}}{\left(  2\pi\right)  ^{3}\upsilon_{\nu,\mathbf{k}_\text{F}^{\prime}}}\sum_{i\omega_{m}}\frac{\gamma_{2}^{Z}}{\gamma_{\nu}^{Z}}\frac{V^{\mu\nu}\left(  \mathbf{k-k^{\prime}},i\omega_{n}-i\omega_{m}\right)  }{\left\vert \omega_{m}Z_{\nu}\left(\mathbf{k^{\prime}},i\omega_{m}\right)  \right\vert }\phi_{\nu}\left(\mathbf{k^{\prime}},i\omega_{m}\right),
\end{equation}
where $\lambda^{\prime}=\gamma_{2}^{Z}\lambda$ is an overall scaling of the eigenvalues. Thus the pairing symmetry may only be modified by the ratio, $\eta_{Z}=\gamma_{2}^{Z}/\gamma_{1}^{Z}$, and the `jungle-gym' Fermi surface becomes dominant when $\eta_{Z}\rightarrow \infty$.

(2) For $\upsilon_{\mu,\mathbf{k}_\text{F}}\rightarrow\gamma_{\mu}^{\upsilon}\upsilon_{\mu,\mathbf{k}_\text{F}}$, the gap equation becomes (for $Z_\mu\gg 1$)
\begin{equation}
\lambda\phi_{\mu}\left(  \mathbf{k},i\omega_{n}\right)    =-C\pi T\sum_{\nu}{\oint_{\text{FS}_{\nu}}}\frac{d\mathbf{k}_{//}^{\prime}}{\left(  2\pi\right)  ^{3}\upsilon_{\nu,\mathbf{k}_\text{F}^{\prime}}}\sum_{i\omega_{m}}\frac{\tilde{V}^{\mu\nu}\left(  \mathbf{k-k^{\prime}},i\omega_{n}-i\omega_{m}\right)  }{\left\vert\sum_{\kappa}\frac{\gamma_{\nu}^{\upsilon}}{\gamma_{\kappa}^{\upsilon}}{\oint_{\text{FS}_{\kappa}}}\frac{d\mathbf{k}_{//}^{\prime\prime}}{\left(  2\pi\right)  ^{3}\upsilon_{\kappa,\mathbf{k}_\text{F}^{\prime\prime}}}P_{\nu\kappa}\left(\mathbf{k}^{\prime}\mathbf{-k^{\prime\prime}},i\omega_{m}\right)  \right\vert}\phi_{\nu}\left(  \mathbf{k^{\prime}},i\omega_{m}\right).
\end{equation}
Similarly, the pairing symmetry may only be modified by the ratio, $\eta_{\upsilon}=\gamma_{2}^{\upsilon}/\gamma_{1}^{\upsilon}$. We find that the `jungle-gym' Fermi surface becomes dominant when $\eta_{\upsilon}\rightarrow \infty$.

\begin{figure}[h]
\centering\includegraphics[width=0.65\textwidth]{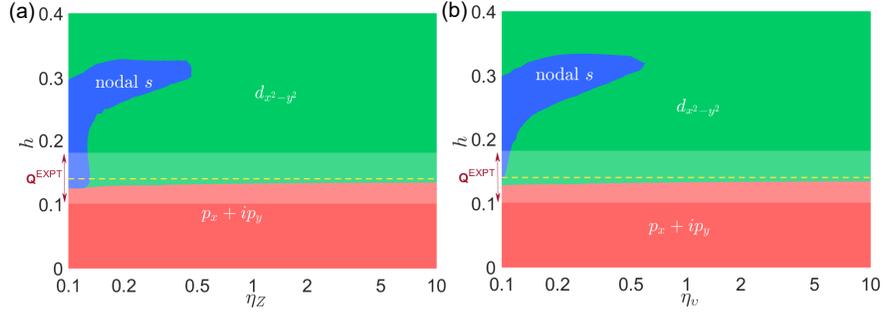}\caption{Superconducting phase diagrams with (a) $(h, \eta_Z)$ and (b) $(h, \eta_\upsilon)$, showing the robustness of the $d$-$p$ phase boundary.  The gray region and dashed line mark the range of $\mathbf{Q}^{\text{EXPT}} = (0.14\pm0.04, 0.14\pm0.04, 0)$.}
\label{figS3}
\end{figure}

The resulting phase diagrams are plotted in Fig.~\ref{figS3}. For both cases, the $d$-$p$ phase boundary remains almost unchanged until $\eta_Z$ or $\eta_{\upsilon}$ becomes as small as 0.1, where a nodal $s$-wave solution, primarily originating from the `doughnut'-like Fermi surface, appears for large $h$. This is way beyond the reasonable range of variations, as our DFT+$U$ calculations are consistent with dHvA measurements and quantum critical fluctuations only lead to an additional enhancement of the mass ratio by roughly 2. We thus conclude that our results are robust against small modification of the mass ratio.

\subsection{IV. Final remarks on the DFT+$U$+QC framework}
We should note that $\eta_{\upsilon}$ and $\eta_Z$ have an opposite effect on the mass ratio, $m_{1}^{\ast}/m_{2}^{\ast}\approx\bar{Z}_{1}\upsilon_{2,\text{F}}/\bar{Z}_{2}\upsilon_{1,\text{F}}$. Our above analyses reveal a crucial difference between the hybridization effect due to background band structures and the renormalization effect due to quantum criticality. The reason is simple: the renormalization function $Z_\mu$ not only affects the effective mass, but also reduces the spectral weight of quasiparticles, which is harmful to the pairing and may become important in dealing with multiband superconductivity. Thus, it is important to distinguish these two effects. As a consequence, it is incorrect to start with a fully renormalized band structure for superconducting calculations in heavy fermion materials. In the absence of a satisfactory theory, the validity of our results stimulates us to think that DFT+$U$+QC might be useful as a more general framework for understanding heavy fermion physics, as long as DFT+$U$ provides the proper topology of the Fermi surfaces and quantum critical interactions provide the major renormalization effect. In some sense, this is equivalent to an effective periodic Anderson-like model with the tight-binding part from band calculations plus additional effective quantum critical interactions. Of course, we cannot exclude the possibility of other important interaction effects, but these may be overcome by extending the framework to include more sophisticated approaches (such as DFT+DMFT) for band calculations, self-energy/vertex corrections or critical fluctuations. From the view of a spin-fermion model, our calculations can only be regarded as the lowest-order approximation that ignores the complicated interplay of fermionic and bosonic degrees of freedom and may need to be revised in the vicinity of the quantum critical point ($\xi\rightarrow\infty$). A phenomenological theory of this type has been used in understanding other correlated systems such as cuprates. It might also be applicable here to provide certain insight from a different angle in understanding both the normal state and superconducting properties of heavy fermion materials.

\end{document}